\documentclass[journal,10pt]{IEEEtran}

\usepackage{amsthm}
\usepackage{amssymb}
\usepackage{mathrsfs}
\usepackage{mathtools}
\usepackage{float}
\usepackage[pdftex]{graphicx}
\usepackage{amsmath}
\usepackage{color}
\usepackage{multirow}
\usepackage{graphicx}
\usepackage{graphicx,float,wrapfig,epstopdf,amsmath}
\usepackage[T1]{fontenc}

\usepackage{balance}
\usepackage[caption=false,font=footnotesize]{subfig}
\usepackage{cite}
\usepackage{hyperref}
\usepackage[table]{xcolor}% http://ctan.org/pkg/xcolor
\definecolor{color1}{RGB}{199,209,232}
\definecolor{color2}{RGB}{230,231,233}
\begin{document}
	%	\the\baselineskip

	%TC:ignore
	
	\title{Implicit Channel Learning for Machine Learning Applications in 6G Wireless Networks
		%		Artificial Intelligence and Channel Estimation: \\ Towards Implicit Channel Learning}
		%	\title{Implicit Wireless Channel Learning \\ in Machine Learning Tasks}
		%	\title{On The Effect of Wireless Channels \\ in Machine Learning Tasks}
	}
	
	\author{\IEEEauthorblockN{Ahmet M. Elbir, Wei Shi, Kumar Vijay Mishra, Anastasios K. Papazafeiropoulos, and  Symeon Chatzinotas
		}
		\thanks{This work was supported in part by the Natural Sciences and Engineering Research Council of Canada (NSERC), Ericsson Canada and the ERC Project AGNOSTIC.}
		\thanks{A. M. Elbir is with the Department of Electrical and Electronics Engineering, Duzce University, Duzce, Turkey, and the SnT at the University of Luxembourg, Luxembourg (e-mail: ahmetmelbir@gmail.com).} 
		\thanks{W. Shi is with the	School of Information Technology, Carleton
			University, Ottawa, Canada (e-mail: wei.shi@carleton.ca).}
		\thanks{K. V. Mishra is with the United States Army Research Laboratory, Adelphi, MD 20783 USA (e-mail: kvm@ieee.org).}
		\thanks{A. K. Papazafeiropoulos is with the CIS Research Group, University of Hertfordshire, Hatfield, U. K., and SnT at the University of Luxembourg, Luxembourg (e-mail: tapapazaf@gmail.com). }
		
		\thanks{S. Chatzinotas is with the SnT at the University of Luxembourg, Luxembourg (e-mail: symeon.chatzinotas@uni.lu). }
		%		\thanks{D. Gunduz is with Information Processing and Communications Lab, Department of Electrical and Electronic Engineering, Imperial College London  (email: d.gunduz@imperial.ac.uk).}
		
	}

	\maketitle

	\begin{abstract}
		With the deployment of the fifth generation (5G) wireless systems gathering momentum across the world, possible technologies for 6G are under active research discussions. 
		In particular, the role of machine learning (ML) in 6G is expected to enhance and aid emerging applications such as virtual and augmented reality, vehicular autonomy, and computer vision. This will result in large segments of wireless data traffic  comprising image, video and speech. The ML algorithms process these for classification/recognition/estimation through the learning models located on cloud servers. This requires wireless transmission of data from edge devices to the cloud server. Channel estimation, handled separately from recognition step, is critical for accurate learning performance. 
		Toward combining the learning for both channel and the ML data, we introduce \textit{implicit channel learning} to perform the ML tasks without estimating the wireless channel. Here, the ML models are trained with channel-corrupted datasets in place of nominal data. Without channel estimation, the proposed approach exhibits approximately $60\%$ improvement in image and speech classification tasks for diverse scenarios such as millimeter wave and IEEE 802.11p vehicular channels.
	\end{abstract}
	\begin{IEEEkeywords}
		Machine learning, channel estimation, artificial intelligence, wireless communications.
	\end{IEEEkeywords}
	
	%TC:endignore

	\section{Introduction}
	%\IEEEPARstart{W}{hile} 
	%	While 
	Lately, the fifth generation (5G) wireless networks are very close to operational deployment. The lessons learned from the previous and ongoing 5G research is paving way for conceptualization of 6G technologies within the wireless communications community~\cite{6GLetaiefJSAC}. Different from the previous generation, the 6G networks are envisioned to enhance and aid emerging applications such as virtual and augmented reality (VAR), autonomous vehicles (AVs) and artificial intelligence (AI) together with ubiquitous connectivity and massive number of devices. In particular, 6G is expected to achieve higher peak data rates ($>100\text{Gb/s}$), lower latency ($<1\text{ms}$) and enhanced reliability ($99.999\%$)~\cite{6GLetaiefJSAC,6G_AIenabled6G}. Furthermore, compared to 5G systems incorporating massive  multiple-input multiple-output (MIMO) configurations, 6G networks will operate at higher frequencies (up to $10$ THz) to exploit high-capacity data transmission with wider bandwidth and ultra-massive (UM) MIMO architectures.

	In order to design the physical layer in 5G networks, recently machine learning (ML) techniques have been introduced for the challenging design problems, e.g., resource management~\cite{dl_WCM,mimoDLChannelModelBeamformingFacebook}, symbol detection~\cite{ganGYeLi_conf}, beamforming~\cite{elbir2019online} and channel estimation~\cite{deepCNN_ChannelEstimation}. While these methods have shown a great potential for system design to deal with data, hardware, and computational complexities, they have not been deployed in the current 5G architectures. Rather, the ML-based techniques are envisioned as the primary candidate to be considered in 6G network design~\cite{6GLetaiefJSAC,6G_AIenabled6G,elbir2021FL4PHY}.
	
	%	 Thus, the ML-oriented  wireless networks can provide a paradigm shift in 6G from ''connected things" to ''connected intelligence" by imparting cognition to the system design~\cite{6GLetaiefJSAC}. 

	%	in such a way that several design aspects such as spectrum sharing, resource management 

\begin{figure*}[t]
		\centering
		{\includegraphics[draft=false,width=1.0\textwidth]{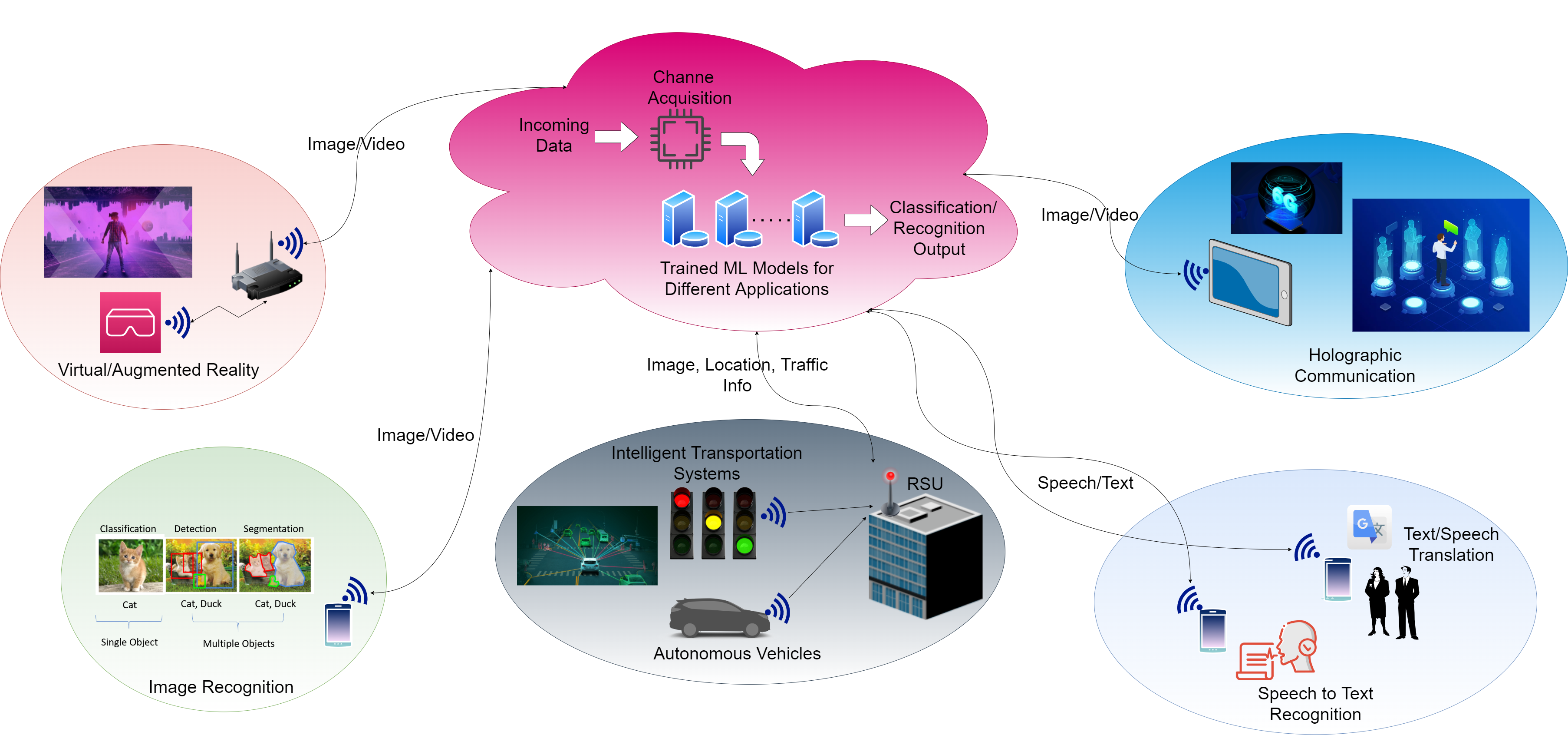} } 
		\caption{Various ML applications for 6G use-cases and the corresponding transmit data types. The edge devices send the ML data to the PS through wireless medium. Once the channel acquisition is performed, the received ML data is input to the learning models. The recognition task is carried out at the PS that sends back the prediction output to the edge devices.
		}
		%			\vspace*{-5mm}
		\label{fig1}
	\end{figure*}

	In the upcoming 6G era, the wireless network will incorporate  massive number of  devices such as mobile phones, connected vehicles, drones, and internet of thing (IoT) devices. Hence, there will be a tremendous increase in the amount of data generated by these devices.   The International telecommunications union (ITU) estimates that the global mobile traffic in 2030 will reach $5016$ exabytes (EB), of which more than $70\%$ will be image or video~\cite{survey_DL_Scalable}. Moreover, AVs are expected to generate approximately 20 TB/day/vehicle data, a part of which is transmitted to the road side infrastructure~\cite{vehicularFederatedMag}. To process and extract useful information from these images/videos, several ML techniques have emerged as a key enabling technology in the field of image classification, face/object/motion detection, target tracking, and speech recognition (Fig.~\ref{fig1}). These applications require huge amount of data to be processed and learned by an ML model for extracting the features from the raw data and provide a ``meaning''.  As a result, most of the processed data in 6G is expected to be generated or processed by ML algorithms.

	%TC:ignore
	%%-----------------------------------------------------
	%%-----------------------------------------------------
	%TC:endignore

	%TC:ignore
	%----------------------------------------------------------------------------------------------------------
	\begin{table*}
		\caption{State-of-the-art for channel estimation techniques
		}
		\label{tableSummary}
		\centering
		\begin{tabular}{p{0.16\textwidth}p{0.21\textwidth}p{0.22\textwidth}p{0.21\textwidth}}
			\hline 
			\hline
			\cellcolor{color2}\bf Channel estimation  &\bf Algorithm \cellcolor{color1} & \cellcolor{color2}\bf Advantages		& \cellcolor{color2}\bf Drawbacks \\
			\hline
			\cellcolor{color2} Model-based for massive MIMO \cite{mimoHybridLeus2}
			& \cellcolor{color1} Codebook-based adaptive compressed sensing 
			& \cellcolor{color2}  Simple implementation especially for single-path channels
			& \cellcolor{color1}Inferior performance because of grid-based channel construction  \\
			\hline
			\cellcolor{color2} Model-based for IEEE 802.11p~\cite{80211pChannelEstimation}
			& \cellcolor{color1}  LS estimation with constructed data pilots 
			& \cellcolor{color2}  Improved representation of the data correlation characteristics
			& \cellcolor{color1}High computational complexity\\
			\hline
			\cellcolor{color2} ML-based for IEEE 802.11p~\cite{vehChanEstDL}
			& \cellcolor{color1} DNN design with spectral-temporal averaging
			& \cellcolor{color2}  Low computational complexity 
			& \cellcolor{color1} Performance is data-dependent and worsens for new data  \\
			\hline
			\cellcolor{color2} ML-based in massive MIMO~\cite{elbir2019online}
			& \cellcolor{color1}  Offline and online deployment with hybrid beamforming
			& \cellcolor{color2}  Implicit via CNN fed with received pilots
			& \cellcolor{color1}  Limited performance at high SNR because of low resolution  \\
			\hline
			\cellcolor{color2}  ML-based in massive MIMO~\cite{deepCNN_ChannelEstimation}
			& \cellcolor{color1}  CoNN design with spatial-frequency-temporal  correlation
			& \cellcolor{color2}  Low complexity  for wideband systems 
			& \cellcolor{color1} Requires full matrix inversion for input design  \\
			\hline
			\cellcolor{color2} Implicit for massive MIMO~\cite{mimoDLChannelModelBeamformingFacebook}
			& \cellcolor{color1}  MLP model with codebook-based beam training
			& \cellcolor{color2}  Bypassing the channel estimation stage for hybrid beamforming
			& \cellcolor{color1}Requires prior, e.g., user locations and environment geometry \\
			\hline
			\cellcolor{color2} Implicit for massive MIMO~\cite{ganGYeLi_conf}
			& \cellcolor{color1}  Conditional GAN model
			& \cellcolor{color2}  Bypassing the channel estimation stage for signal detection
			& \cellcolor{color1}Requires prior, e.g., received pilot signals \\
			\hline
			\cellcolor{color2} Implicit for IRS-assisted massive MIMO~\cite{implicitCE_IRS}
			& \cellcolor{color1}  Graph neural network architecture for beamformer design
			& \cellcolor{color2} Bypassing the channel estimation stage to obtain IRS beamformers 
			& \cellcolor{color1} Requires prior, e.g., user locations \\
			\hline
			\cellcolor{color2} Implicit for ISAC~\cite{implicitCE_ISAC}
			& \cellcolor{color1}  Path loss and Doppler shift estimation  via grid search
			& \cellcolor{color2}  Bypassing the channel estimation stage for transmitter design
			& \cellcolor{color1}Requires prior, e.g., location and speed of the vehicles  \\
			\hline
			\hline
		\end{tabular}
	\end{table*}
	%------------------------------------------------------------
	%TC:endignore

	The performance of the ML models depends on the data collected from the edge devices in the network. However the wireless channel corrupts ML data (image/video/speech) during transmission and reduces the inference performance~\cite{ganGYeLi_conf}. Thus, wireless channel acquisition is a crucial task in the ML applications. Furthermore, the 6G requirements, e.g., ultra-low latency and high mobility, make the channel acquisition process even more challenging~\cite{6GLetaiefJSAC}. To this end, several model-based~\cite{mimoHybridLeus2,80211pChannelEstimation} and ML-based~\cite{deepCNN_ChannelEstimation,elbir2019online,vehChanEstDL} channel state information (CSI) estimation techniques have been proposed for various communications standards in cellular (5G millimeter wave (mmWave)) and vehicular (IEEE 802.11p) networks. The model-based techniques have high energy and time consumption because they usually require tackling a high-dimensional optimization problem, especially for UM-MIMO configuration.	On the other hand, the ML-based methods are data-driven, learn the features of the raw data, and provide robustness against noise and channel corruptions. Thus, it is instructive to leverage ML to address the uncertainty in channel estimates. However, implementation of ML techniques necessitates the task-oriented design of multiple ML models, each of which is dedicated to a different layer in the Open System Interconnection (OSI) communications model~\cite{6G_AIenabled6G}. For instance, separate ML models for physical (e.g., channel estimation) and application (e.g., image recognition) layer tasks should be devised. 
	
	%In this article, we are 
	Motivated by the aforementioned two 6G facets --- ML-related data and ML-based network design --- we introduce \textit{implicit channel learning} approach by combining the learning for both wireless channel and ML-related data.  To this end, an ML model is trained with the dataset carrying the accompanied wireless channel such as path loss and correlation. Thus, the trained model implicitly learns the channel characteristics and makes accurate predictions even if the ML data are corrupted by the wireless channel. While there are a few studies on the implicit estimation of wireless channel for symbol detection~\cite{ganGYeLi_conf} and active/passive beamforming~\cite{implicitCE_IRS,implicitCE_ISAC,mimoDLChannelModelBeamformingFacebook}, these methods require either \textit{a priori} information such as user locations, or received pilot signals along with the optimization of the remaining system parameters, e.g., beamformers (see Table~\ref{tableSummary}). Unlike these prior works, our proposed approach addresses both channel learning and application layer learning  tasks through a single ML model, which is fed with the channel corrupted image/speech data. Thus, it is helpful for future 6G networks that are likely to heavily utilize ML-related data.% and reduce the channel estimation complexity.

	Our work is connected with the rich body of signal processing literature on classic problems of \textit{blind channel estimation}, \textit{blind equalization}, and \textit{blind decoding}, where the channel is unknown and not estimated \textit{a priori} while decoding the communications messages. Instead, our technique predicts the classification/recognition information of the transmitted data.
	
	In the next section, we discuss common applications of ML in wireless communications. Next, we survey the state-of-the-art in channel estimation for various network architectures such as cellular and vehicular networks (see Table~\ref{tableSummary}). In this context, we introduce our {implicit channel learning} approach and validate its performance on image and speech classification tasks under the effect of various wireless standards. 
	Finally, we identify the major design challenges in realizing implicit channel learning and highlight some future research directions.

	\section{ML Applications in Wireless Networks}
	%In this section, we first discuss  how the ML-based CV tasks such as image, video, speech, and text classification, are handled by the wireless communications channel. Next, we survey the ML techniques for channel estimation, along with the design challenges therein. 

	%	The ML techniques are data-driven, learn the features of the raw data and provide robustness against noise and channel corruptions. % arising from external factors such as noise degradation and wireless channel. 
	%	Thus, it is natural to leverage ML for the uncertainty in channel estimates. %In fact, several ML-based channel estimation techniques have been proposed recently. 
	%
	%	In \cite{deepCNN_ChannelEstimation}, a deep CNN was designed for channel estimation in broadband massive multiple-output multiple input (mMIMO) systems. In order to provide adaptation in the propagation environment, ML-based online/offline channel estimation techniques were introduced in~\cite{elbir2019online}. 
	
	%	The implementation of these techniques necessitates the task-oriented design of multiple ML models, each of which is dedicated to a different layer in the OSI (Open Systems Interconnection) communications model. For instance, separate ML models for channel estimation in the physical layer and AI tasks in the application layer should be devised. 

	In many contemporary ML applications, training and prediction of the ML model are carried out at a central parameter server (PS) of the network, whereas the data are generated at the network edge comprising devices like mobile phones, vehicles, and IoT sensors. These ML models are composed of huge number of parameters.  For instance,  well-known ML models such as AlexNet, VGG (both for image classification), and GPT-3 (text recognition/translation) are comprised of $60$ million ($240$ MB), $136$ million ($552$ MB), and $170$ billion ($680$ GB) learnable parameters, respectively~\cite{survey_DL_Scalable,elbir2021FL4PHY}. Storing these huge models in the edge devices is costly and inefficient. %due to the memory constraints. 
	Instead, these models are stored at cloud PSs. The training stage is managed offline using a pre-collected training dataset. But the prediction-stage classification/recognition demands transmission of the generated image/video/speech to the PS because the ML models run short of storage at the edge devices. Once the prediction is performed at the PS, its result is sent back to the edge device (Fig.~\ref{fig1}). For example, Google's image recognition technology Lens transmits captured images via a smart phone to the PS, where a pre-trained convolutional neural network (CNN) used for recognition/classification/translation brings up the information related to the objects in the captured image. Similarly, the AVs have on-board pre-trained learning models but still perform data transmission (reception) to (from) the cloud platforms for map generation, path planning, and forecasting~\cite{vehicularFederatedMag}.

	%	   In Figure~\ref{fig1}, this process is illustrated for both V2V and LTE/NR communications channels. \textcolor{blue}{already referenced Fig.1 before} While the mobile phones use LTE/NR channel, vehicles usually form a clustered networks to lower the usage of cellular infrastructure, thus they first communicate over the IEEE 802.11p V2V link and the member of the cluster that is closest to the base station (BS), conveys the information to the BS via vehicular-to-infrastructure (V2I) link~\cite{80211pChannelEstimation}. At the BS, the LTE (5G NR) infrastructure  eNodeB (gNodeB) performs several signal processing operations such as channel equalization, then the information is conveyed to the PS via evolved packet core (EPC), which consists of a server gateway (SGW) and packet data network gateways (PGWs)~\cite{vehChanEstDL}. At the PS, the channel equalized CV data is fed to the ML model, which makes the prediction/recognition depending on the application, such as next-word prediction and image/video/face/speech recognition, and then sends back the output to the network edge. \textcolor{red}{so much repetitive information here}

	%TC:ignore
	%%-----------------------------------------------------
	\begin{figure}[t]
		\centering
		{\includegraphics[draft=false,width=\columnwidth]{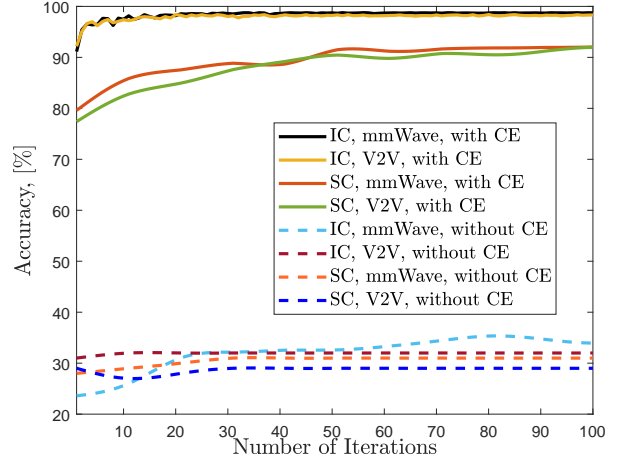} } 
		\caption{Impact of channel estimation (CE) for image (IC) and speech classification (SC) performance on the validation datasets. The details for the datasets and the learning models are given in Table~\ref{tableImage}.
		}
		%			\vspace*{-5mm}
		\label{fig_acc1}
	\end{figure}
	%%-----------------------------------------------------
	%TC:endignore

	%\subsection{The Effect of Wireless Channel}
	Fig.~\ref{fig_acc1} illustrates the impact of channel estimation on ML applications of image and speech classification for mmWave~\cite{mimoHybridLeus2} and vehicle-to-vehicle (V2V)~\cite{80211pChannelEstimation} line-of-sight (LoS) links. We used MNIST and Speech Command datasets for image and speech classification tasks, respectively, each of which has $10$ classes~\cite{survey_DL_Scalable}. The image dataset includes the black-and-white handwritten digits whereas the speech dataset is composed of the spectrogram of audio signals. %Further details regarding datasets and ML model for both tasks are presented in Table~\ref{tableImage}. 
	During training, \textit{clean} (with equalized channel) datasets are used for both ML models while two different validation datasets are prepared with (\textit{clean}) and without (\textit{corrupted}) channel estimation to observe its effect on learning accuracy. We observe that the classification performance degrades significantly if the images/spectrograms are corrupted by the wireless channel without equalization for both tasks. This shows the significance of the channel acquisition in ML applications. 

	\begin{figure*}[t]
		\centering
		{\includegraphics[draft=false,width=.9\textwidth]{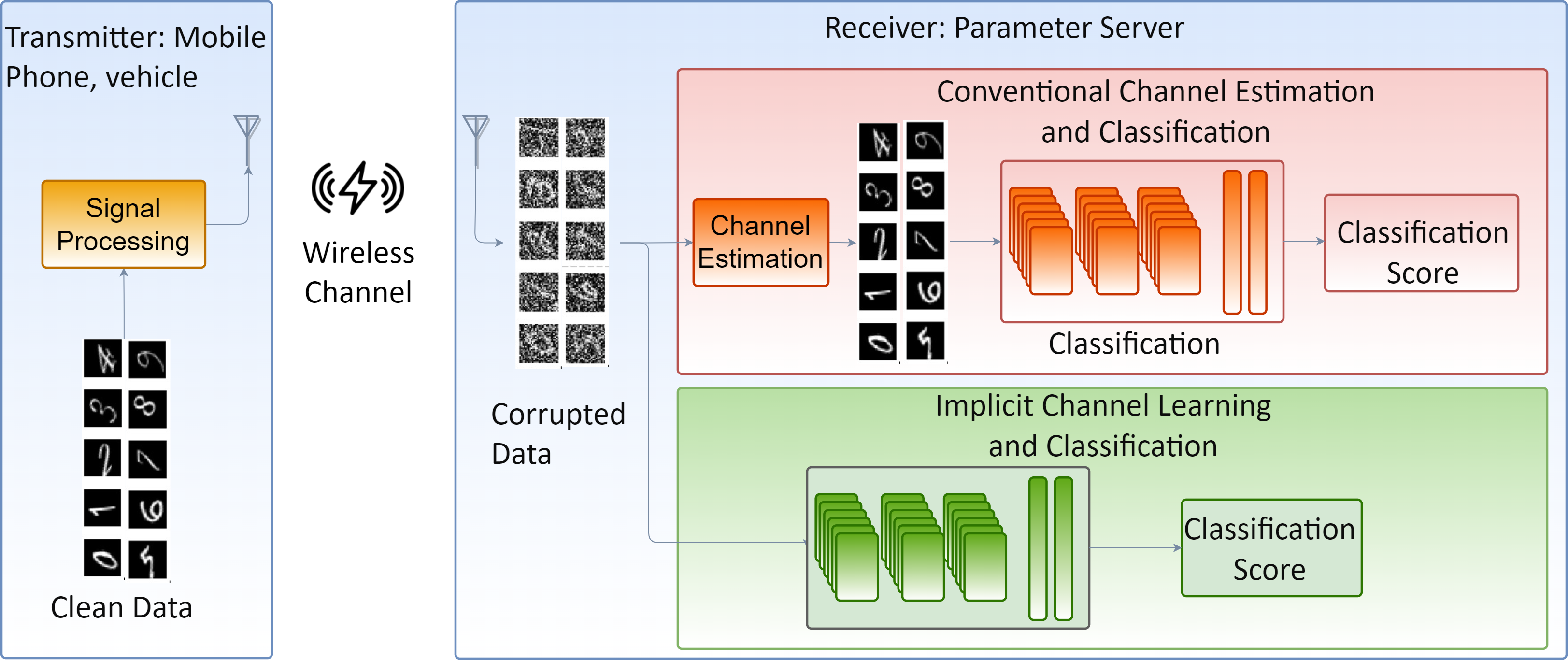} } 
		\caption{Joint image classification and implicit wireless channel learning, wherein the clean data are processed and transmitted through the wireless channel to the PS. Next, the reconstructed channel-corrupted data are fed to the learning model for classification.
		}
		%			\vspace*{-5mm}
		\label{fig_Learning}
	\end{figure*}
	%%-----------------------------------------------------
	%TC:endignore

	\section{Channel Estimation Techniques}	
	
	%In this section, we discuss the various channel estimation techniques (see Table~\ref{tableSummary}) and their advantages/drawbacks.
	The model-based signal processing techniques need accurate mathematical modeling of the transmitted/received signals. However, to address the uncertainties and non-linearities imposed by channel equalization and hardware impairments, model-free ML techniques have become common in wireless communications~\cite{deepCNN_ChannelEstimation,elbir2021FL4PHY}.

	\subsection{Model-based approaches}
	%The model-based signal processing techniques need accurate mathematical modeling of the transmitted/received signals. 
	%In this regard, 
	Channel estimation is an essential task for reliable communication~\cite{mimoHybridLeus2}. However, it is more challenging in %Compared to conventional cellular systems in 5G, channel estimation is a more challenging task in  
	6G architectures, wherein number of antennas in UM arrays is exceedingly huge~\cite{mimoHybridLeus2,6G_AIenabled6G}. Furthermore, the multi-hop communications frameworks, such as vehicular networks~\cite{vehicularFederatedMag,vehChanEstDL} and intelligent reflecting surface (IRS)-empowered systems~\cite{implicitCE_IRS,elbir2021FL4PHY} make channel estimation even more demanding. %  multiple channel links. 
	In particular, vehicular networks have highly dynamic channels arising from rapid vehicular mobility (up to $150$ km/h  in 6G). Then, variation in weather conditions (resulting in a path
	loss of \textasciitilde $4$ dB over $0.1$-$0.3$ THz) causes frequent drop-outs and hand-overs~\cite{vehicularFederatedMag,80211pChannelEstimation}.  As a result, there exists an inherit uncertainty stemming from the dynamics of the wireless channel in both network architectures. %To address the uncertainties and non-linearities imposed by channel equalization and hardware impairments, model-free ML techniques have become common in wireless communications~\cite{deepCNN_ChannelEstimation,elbir2021FL4PHY}. 
	
	\subsection{ML-based approaches}
	Massive MIMO channel estimation via ML is investigated in~\cite{deepCNN_ChannelEstimation}, where a convolutional-only neural network (CoNN) is designed. The input of the CoNN is the tentative channel matrix that is computed via least squares (LS) method and the output is the channel matrix from multiple subcarriers. The CoNN exhibits improved channel estimation accuracy in terms on the mean-squared error (MSE). But its input data requires matrix inversion, which is computationally inefficient for large antenna systems. To reduce the complexity in preparing the input data,  \cite{elbir2019online} devised a CNN approach for channel estimation by feeding the CNN with the received pilot signals. This approach not only eliminates matrix inversion, but also estimates the full CSI (in the scale of the number of antennas) from reduced-rank input data (in the scale of number of pilot signals, fewer than the number of antennas). This is more applicable to massive MIMO systems, where low dimensional radio-frequency (RF) precoders limit the dimension of the received data~\cite{mimoHybridLeus2}. Nevertheless, training %of the ML model 
	in~\cite{deepCNN_ChannelEstimation} and \cite{elbir2019online} is based on centralized schemes, for which the training data needs to be transmitted from the edge devices (mobile phones) to the PS. This results in huge communications overhead arising from large sizes of training datasets. This may be mitigated by employing federated learning (FL)~\cite{elbir2021FL4PHY,vehicularFederatedMag}, wherein CNN parameters are computed at edge devices based on their local datasets not shared with the PS; instead, only the model parameters/updates are transmitted to the PS for model aggregation.
	
	%Apart from the aforementioned ML works in cellular networks, 
	For vehicular networks, ML-based channel estimation in~\cite{vehChanEstDL} used a multi-layer perceptron (MLP) architecture to enhance the tentative channel estimate (TCE) obtained via spectral-temporal averaging (STA). This incorporated time and frequency correlation ratios of two successive orthogonal frequency division multiplexing (OFDM) symbols. The input to MLP was STA-based TCE while the output was the true channel vector labels.

	%TC:ignore
	%----------------------------------------------------------------------------------------------------------
	\begin{table*}[ht]
		\caption{Image (left) and speech (right) classification performance under different channel conditions.
			%Results are Given for Clean and Corrupted Test Datasets, respectively.
		}
		% \hfill%
		%\ttabbox%
		\label{tableImage}
		\centering
		\begin{tabular}{ cc }   % top level tables, with 2 columns
			\cellcolor{color1}
			\begin{tabular}
				{p{0.05\textwidth}p{0.05\textwidth}p{0.05\textwidth}p{0.05\textwidth}p{0.05\textwidth}p{0.05\textwidth}}
				\hline 
				\hline
				\multicolumn{6}{c}{\cellcolor{color2} \parbox{7.5cm}{ {\bf Learning Model:} CNN with two convolutional layers (128@$5\times 5$ and 128@ $3\times 3$)  and a single fully connected layer ($128$ units)\\
						{\bf Dataset:} MNIST Handwritten Digits Dataset ($28\times 28\times 60,000$)
						\\ {\bf Classes:} $\{0,1,2,3,4,5,6,7,8,9\}$ \\ } }\\
				\hline \hline
				\multicolumn{2}{p{0.125\textwidth}}{\cellcolor{color1}\centering  Training Data: \par  \texttt{D1} \par Test Data: \par  Clean - {\bf Corrupted}}%,\par
				&\multicolumn{2}{p{0.125\textwidth}}{\cellcolor{color2} \centering  Training Data: \par \texttt{D2} \par Test Data: \par  Clean - {\bf Corrupted}}%, \par Test Data: Clean 
				&\multicolumn{2}{p{0.125\textwidth}}{\cellcolor{color1}\centering   Training Data:\par  \texttt{D3} \par Test Data: \par  Clean - {\bf Corrupted}}%, \par Test Data: Clean 
				\tabularnewline
				\hline \hline
				\multicolumn{6}{c}{\cellcolor{color2}\bf mmWave}\\
				\hline
				\cellcolor{color1}$98.6\% $&	\cellcolor{color1}\centering$ \bf34.4\%$
				&\cellcolor{color2}$93.3\%$&\cellcolor{color2}\centering$\bf97.1\%$\centering
				&\cellcolor{color1}\centering $95.1\%$&	\cellcolor{color1}\centering$\bf  97.0\%$
				\tabularnewline
				\hline		\hline 
				\multicolumn{6}{c}{\cellcolor{color2}\bf V2V - Rural LoS}\\
				\hline
				\cellcolor{color1}$98.6\% $&	\cellcolor{color1}\centering$ \bf32.1\%$
				&\cellcolor{color2}$92.0\% $&\cellcolor{color2}\centering$ \bf95.1\%$\centering
				&\cellcolor{color1}$93.3\% $&	\cellcolor{color1}\centering$\bf 94.3\%$\centering
				\tabularnewline
				\hline 		\hline 
				\multicolumn{6}{c}{\cellcolor{color2}\bf V2V - Urban  LoS}\\
				\hline 
				\cellcolor{color1}$98.6 \%$&	\cellcolor{color1}\centering$\bf 27.7\%$\centering
				&\cellcolor{color2}$97.8\%$&\cellcolor{color2}\centering$\bf87.9\%$\centering
				&\cellcolor{color1}$98.1\%$&	\cellcolor{color1}\centering$\bf 86.2\%$\centering
				\tabularnewline
				\hline 		\hline 
				\multicolumn{6}{c}{\cellcolor{color2}\bf V2V - Urban nLoS}\\
				\hline
				\cellcolor{color1}$98.6\%$&	\cellcolor{color1}\centering$ \bf15.2\%$\centering
				&\cellcolor{color2}$97.1\%$&\cellcolor{color2}\centering$\bf87.7\%$\centering
				&\cellcolor{color1}$98.0\%$&	\cellcolor{color1}\centering$\bf 85.8\%$\centering
				\tabularnewline
				\hline 		\hline 
				\multicolumn{6}{c}{\cellcolor{color2} \bf V2V - Highway LoS}\\
				\hline 
				\cellcolor{color1}$98.7\% $&	\cellcolor{color1}\centering$\bf 30.4\%$\centering
				&\cellcolor{color2}$97.9\%$&\cellcolor{color2}\centering$\bf90.3\%$\centering
				&\cellcolor{color1}$98.3\%$&	\cellcolor{color1}\centering$\bf 88.4\%$\centering
				\tabularnewline
				\hline 		\hline 
				\multicolumn{6}{c}{\cellcolor{color2}\bf V2V - Highway nLoS}\\
				\hline 
				\cellcolor{color1}$98.7\% $&	\cellcolor{color1}\centering$\bf 13.5\%$\centering
				&\cellcolor{color2}$96.9\%$&\cellcolor{color2}\centering$\bf97.7\%$\centering
				&\cellcolor{color1}$98.2\%$&	\cellcolor{color1}\centering$ \bf85.6\%$\centering
				\tabularnewline
				\hline		\hline 
				\multicolumn{6}{c}{\cellcolor{color2}\bf mmWave-V2V - Rural LoS}\\
				\hline
				\cellcolor{color1}$98.3\%$&	\cellcolor{color1}\centering$\bf 27.5\%$\centering
				&\cellcolor{color2}$90.4\% $&\cellcolor{color2}\centering$\bf92.2\%$\centering
				&\cellcolor{color1}$91.1\% $&	\cellcolor{color1}\centering$\bf 91.9\%$\centering
				\tabularnewline
				\hline 		\hline 
				\multicolumn{6}{c}{\cellcolor{color2}\bf mmWave-V2V - Urban  LoS}\\
				\hline 
				\cellcolor{color1}$98.2 \%$&	\cellcolor{color1}\centering$\bf 31.0\%$\centering
				&\cellcolor{color2}$90.5\%$&\cellcolor{color2}\centering$\bf92.8\%$\centering
				&\cellcolor{color1}$90.3\%$&	\cellcolor{color1}\centering$ \bf91.1\%$\centering
				\tabularnewline
				\hline 		\hline 
				\multicolumn{6}{c}{\cellcolor{color2}\bf mmWave-V2V - Urban nLoS}\\
				\hline
				\cellcolor{color1}$97.7\% $&	\cellcolor{color1}\centering$\bf 12.5\%$\centering
				&\cellcolor{color2}$83.8\%$&\cellcolor{color2}\centering$\bf88.1\%$\centering
				&\cellcolor{color1}$86.1\%$&	\cellcolor{color1}\centering$\bf 87.4\%$\centering
				\tabularnewline
				\hline 		\hline 
				\multicolumn{6}{c}{\cellcolor{color2}\bf mmWave-V2V - Highway LoS}\\
				\hline 
				\cellcolor{color1}$98.4\%$&	\cellcolor{color1}\centering$\bf 32.0\%$\centering
				&\cellcolor{color2}$90.4\%$&\cellcolor{color2}\centering$\bf90.3\%$\centering
				&\cellcolor{color1}$89.0\%$&	\cellcolor{color1}\centering$\bf 89.8\%$\centering
				\tabularnewline
				\hline 		\hline 
				\multicolumn{6}{c}{\cellcolor{color2}\bf mmWave-V2V - Highway nLoS}\\
				\hline 
				\cellcolor{color1}$97.1\% $&	\cellcolor{color1}\centering$\bf 17.6\%$\centering
				&\cellcolor{color2}$84.4\%$&	\cellcolor{color2}\centering$\bf88.1\%$\centering
				&\cellcolor{color1}$86.5\%$&	\cellcolor{color1}\centering$\bf 87.2\%$\centering
				\tabularnewline
			\end{tabular} &
			%\end{table*}
			%%------------------------------------------------------------
			%\begin{table}[t!]
			%	\caption{Speech  Classification Performance  in Various Channel Conditions. 
			%		%Results are Given for Clean and Corrupted Test Datasets, respectively.
			%	}
			%	\label{tableSpeech}
			%	\centering
			\cellcolor{color1}
			\begin{tabular}
				{p{0.05\textwidth}p{0.05\textwidth}p{0.05\textwidth}p{0.05\textwidth}p{0.05\textwidth}p{0.05\textwidth}}
				\hline 
				\hline
				\multicolumn{6}{c}{\cellcolor{color2} \parbox{7.5cm}{ {\bf Learning Model:} CNN with four convolutional layers (16@$5\times 5$, 16@$3\times 3$, 16@$3\times 3$ and 16@$3\times 3$)  and a single fully connected layer ($128$ units)  \\ 
						{\bf  Dataset:} Speech Command Dataset ($98\times 50\times 28,370$) 
						\\ {\bf Classes:} $\{\mathrm{yes},
						\mathrm{no}
						, \mathrm{up}
						, \mathrm{down}
						, \mathrm{left}
						, \mathrm{right}
						, \mathrm{on}
						, \mathrm{off}
						,\mathrm{stop}
						,\mathrm{go}
						\}$  }} \\
				\hline \hline
				\multicolumn{2}{p{0.125\textwidth}}{\cellcolor{color1}\centering  Training Data: \par \texttt{D1} \par Test Data: \par  Clean - {\bf Corrupted}}%,\par
				&\multicolumn{2}{p{0.125\textwidth}}{\cellcolor{color2} \centering  Training Data: \par \texttt{D2} \par Test Data: \par  Clean - {\bf Corrupted}}%, \par Test Data: Clean 
				&\multicolumn{2}{p{0.125\textwidth}}{\cellcolor{color1}\centering   Training Data:\par  \texttt{D3}\par Test Data: \par  Clean - {\bf Corrupted}}%, \par Test Data: Clean 
				\tabularnewline
				\hline 		\hline 
				\multicolumn{6}{c}{\cellcolor{color2}\bf mmWave}\\
				\hline
				\cellcolor{color1}$91.5\% $&	\cellcolor{color1}\centering$\bf 31.6\%$
				&\cellcolor{color2}$78.6\%$&\cellcolor{color2}\centering$\bf88.3\%$\centering
				&\cellcolor{color1}\centering $81.8\%$&	\cellcolor{color1}\centering$\bf  87.1\%$
				\tabularnewline
				\hline		\hline 
				\multicolumn{6}{c}{\cellcolor{color2}\bf V2V - Rural LoS}\\
				\hline
				\cellcolor{color1}$91.5\% $&	\cellcolor{color1}\centering$\bf 29.6\%$
				&\cellcolor{color2} $ 77.1\% $&\cellcolor{color2}\centering$ \bf87.5\%$\centering
				&\cellcolor{color1}$79.8\% $&	\cellcolor{color1}\centering$\bf 86.3\%$\centering
				\tabularnewline
				\hline 		\hline 
				\multicolumn{6}{c}{\cellcolor{color2}\bf V2V - Urban  LoS}\\
				\hline 
				\cellcolor{color1}$91.6 \%$&	\cellcolor{color1}\centering$\bf 27.7\%$\centering
				&\cellcolor{color2}$77.5\%$&\cellcolor{color2}\centering$\bf85.6\%$\centering
				&\cellcolor{color1}$80.0\%$&	\cellcolor{color1}\centering$ \bf84.7\%$\centering
				\tabularnewline
				\hline 		\hline 
				\multicolumn{6}{c}{\cellcolor{color2}\bf V2V - Urban nLoS}\\
				\hline
				\cellcolor{color1}$91.5\%$&	\cellcolor{color1}\centering$\bf 15.2\%$\centering
				&\cellcolor{color2}$76.7\%$&\cellcolor{color2}\centering$\bf85.5\%$\centering
				&\cellcolor{color1}$78.4\%$&	\cellcolor{color1}\centering$\bf 84.7\%$\centering
				\tabularnewline
				\hline 		\hline 
				\multicolumn{6}{c}{\cellcolor{color2}\bf V2V - Highway LoS}\\
				\hline 
				\cellcolor{color1}$91.5\% $&	\cellcolor{color1}\centering$\bf 27.2\%$\centering
				&\cellcolor{color2}$79.0\%$&\cellcolor{color2}\centering$\bf87.0\%$\centering
				&\cellcolor{color1}$80.2\%$&	\cellcolor{color1}\centering$\bf 86.4\%$\centering
				\tabularnewline
				\hline 		\hline 
				\multicolumn{6}{c}{\cellcolor{color2}\bf V2V - Highway nLoS}\\
				\hline 
				\cellcolor{color1}$91.5\% $&	\cellcolor{color1}\centering$\bf 13.5\%$\centering
				&\cellcolor{color2}$75.3\%$&\cellcolor{color2}\centering$\bf85.1\%$\centering
				&\cellcolor{color1}$77.1\%$&	\cellcolor{color1}\centering$\bf 85.1\%$\centering
				\tabularnewline
				\hline		\hline 
				\multicolumn{6}{c}{\cellcolor{color2}\bf mmWave-V2V - Rural LoS}\\
				\hline
				\cellcolor{color1}$91.5\%$&	\cellcolor{color1}\centering$\bf 27.3\%$\centering
				&\cellcolor{color2}$78.7\% $&\cellcolor{color2}\centering$\bf85.1\%$\centering
				&\cellcolor{color1}$82.6\% $&	\cellcolor{color1}\centering$\bf 84.0\%$\centering
				\tabularnewline
				\hline 		\hline 
				\multicolumn{6}{c}{\cellcolor{color2}\bf mmWave-V2V - Urban  LoS}\\
				\hline 
				\cellcolor{color1}$91.6 \%$&	\cellcolor{color1}\centering$\bf 25.8\%$\centering
				&\cellcolor{color2}$77.8\%$&\cellcolor{color2}\centering$\bf84.2\%$\centering
				&\cellcolor{color1}$82.3\%$&	\cellcolor{color1}\centering$\bf 83.4\%$\centering
				\tabularnewline
				\hline 		\hline 
				\multicolumn{6}{c}{\cellcolor{color2}\bf mmWave-V2V - Urban nLoS}\\
				\hline
				\cellcolor{color1}$90.3\% $&	\cellcolor{color1}\centering$\bf 14.1\%$\centering
				&\cellcolor{color2}$76.2\%$&\cellcolor{color2}\centering$\bf83.7\%$\centering
				&\cellcolor{color1}$81.5\%$&	\cellcolor{color1}\centering$\bf 82.5\%$\centering
				\tabularnewline
				\hline 		\hline 
				\multicolumn{6}{c}{\cellcolor{color2}\bf mmWave-V2V - Highway LoS}\\
				\hline 
				\cellcolor{color1}$91.4\%$&	\cellcolor{color1}\centering$\bf 26.3\%$\centering
				&\cellcolor{color2}$78.2\%$&\cellcolor{color2}\centering$\bf86.5\%$\centering
				&\cellcolor{color1}$82.5\%$&	\cellcolor{color1}\centering$\bf 85.1\%$\centering
				\tabularnewline
				\hline 		\hline 
				\multicolumn{6}{c}{\cellcolor{color2}\bf mmWave-V2V - Highway nLoS}\\
				\hline 
				\cellcolor{color1}$91.4\% $&	\cellcolor{color1}\centering$\bf 12.4\%$\centering
				&\cellcolor{color2}$72.1\%$&	\cellcolor{color2}\centering$\bf83.4\%$\centering
				&\cellcolor{color1}$81.8\%$&	\cellcolor{color1}\centering$\bf 82.0\%$\centering
				\tabularnewline
			\end{tabular}\\
		\end{tabular}
	\end{table*}
	%------------------------------------------------------------
	%TC:endignore
	
	\subsection{Shortcomings of ML-based channel estimation}
	%Despite the aforementioned advantages, 
	The performance of ML-based channel estimation techniques is upper-bounded by the accuracy of the labeling algorithm, which is usually model-based% approach employed to obtain the channel estimate to be used in the training stage
	~\cite{vehChanEstDL,deepCNN_ChannelEstimation}. These algorithms generally rely on the specific mathematical model of the wireless channel and matrix processing steps such as singular value decomposition (SVD)~\cite{mimoDLChannelModelBeamformingFacebook}, matrix inversion~\cite{deepCNN_ChannelEstimation}, and covariance computation~\cite{mimoHybridLeus2}. As a result, labeling remains one of the major challenges in ML-based channel estimation. While physical layer applications such as beamformer design or resource allocation handle labeling via optimization techniques, the ML-based channel estimation methods~\cite{deepCNN_ChannelEstimation,elbir2019online,vehChanEstDL} assume the label as true channel data. When imperfect channel data are designated as label, the estimation accuracy degrades in the MSE sense~\cite{elbir2021FL4PHY}. Obviously, the choice of label-generation algorithm is critical in ensuring the ML performance. 
	
	%In addition, an ML model should be dedicated for channel estimation task whereas the effect of the channel can be mitigated in the CV tasks as will be described extensively in the following section.

	\section{Implicit Channel Learning in ML Tasks}
	Instead of designing dedicated learning models for both channel estimation and ML applications, both tasks could be performed jointly. Here, the ML model is trained on a training dataset while accounting for the imperfections or changes in the wireless channel. This allows the model to learn the corruptions in the ML data and perform the recognition tasks without performing channel estimation, leading to an overall reduction in the computational complexity and channel overhead.

	%\subsection{Performance Results}

	The joint learning of the channel and the ML data requires preparing the training dataset for various channel conditions so that the ML model can extract the pattern in the input data and be robust against the corruptions of the wireless channel. Fig.~\ref{fig_Learning} illustrates the processing chain of training data generation. Here, we generated three datasets each for both image and speech classification. The first \texttt{D1} was \textit{clean} while the second (\texttt{D2}) contained the corrupted data. Third set (\texttt{D3}) is a mix of $50\%$ clean and $50\%$ corrupted data. During data generation, the wireless channel statistics were randomly changed for each image transmission and the signal-to-noise-ratio (SNR) was set to $15$ dB. Each corrupted dataset included $100$ distinct corrupted copies of the clean dataset to provide robustness against imperfections. As a result, the number of samples in these three training datasets were $600,000$, $6,000,000$ and $3,300,000$ ($283,700$, $2,837,000$ and $1,560,350$) for MNIST (Speech Command) dataset, respectively. For the MNIST (Speech Command) dataset, there were two test datasets, each of which had a size of $10,000$ ($4,000$) and they were generated separately from the training data.
	%, and the first one is clean while the second one is corrupted.
	
	Table~\ref{tableImage} shows image and speech classification performance of the learning models for various channel conditions such as mmWave, V2V, and multi-hop mmWave-V2V channels. The V2V channel was tested for multiple delay profiles corresponding to different scenarios, such as Rural LoS, Urban LoS/nLoS (non-LoS), and Highway LoS/nLoS~\cite{80211pChannelEstimation,vehChanEstDL}. We observe that the learning performance was poor if there was a mismatch between the training and test datasets. In particular, the model trained on \texttt{D1} was unable to recognize the corrupted dataset, which includes the channel effects. On the other hand, the models trained on \texttt{D2}  and \texttt{D2} were able to learn the corrupted ML data (image or speech) while implicitly learning the channel characteristics. Compared to the case conducted without channel estimation, they exhibit approximately $60\%$ improvement for both image and speech classification. Furthermore, the model with \texttt{D2} provided slightly higher accuracy than the one possessing \texttt{D3} for the corrupted test data while the latter had a slight performance loss (approximately $1\%$) for clean test dataset. This is because the size of \texttt{D3} is smaller, including both clean and corrupted data. Nevertheless, \texttt{D3} presents satisfactory performance for both tasks with a smaller dataset. Consequently, this suggests that constructing the half of the dataset with channel effects can yield a reliable recognition performance as well as implicitly learning the channel effect on the ML data. 
	
	%TC:ignore
	%----------------------------------------------------------------------------------------------------------
	\begin{table*}[t]
		\caption{Implicit learning applications}
		\label{tableMultiTasks}
		\centering
		\begin{tabular}{p{0.20\textwidth} p{0.35\textwidth} p{0.18\textwidth}  p{0.18\textwidth} }
			\hline
			\hline
			\textbf{Application}\cellcolor{color1} &\textbf{ Loss function}
			\cellcolor{color2}  
			& \textbf{Input }
			\cellcolor{color1} &\textbf{ Output}
			\cellcolor{color2}  \\
			Implicit channel learning\cellcolor{color2} & Cross-entropy cost for channel corrupted data\cellcolor{color1} & Channel corrupted data\cellcolor{color2} &Classification score \cellcolor{color1}  \\
			\hline
			Implicit channel and beamformer learning\cellcolor{color1} & Cross-entropy cost for channel- and beamformer- corrupted data\cellcolor{color2} & Channel- and beamformer- corrupted data \cellcolor{color1} & Classification score\cellcolor{color2} \\
			\hline
			Implicit channel learning and semantic communication\cellcolor{color2} & Cross-entropy cost for channel corrupted data and mutual information \cellcolor{color1} &A sentence with semantic encoding \cellcolor{color2} & Recovered sentence\cellcolor{color1} \\
			\hline
			Implicit channel learning for beamformer design\cellcolor{color2} & MSE between label and predicted beamformers \cellcolor{color1} & Received pilot signals \cellcolor{color2} & Beamformer weights \cellcolor{color1} \\
			\hline
			Implicit channel learning for user localization\cellcolor{color1} & MSE between label and predicted locations \cellcolor{color2} &Received pilot signals \cellcolor{color1} & User locations/directions \cellcolor{color2} \\
			\hline
			Implicit channel learning for antenna selection\cellcolor{color2} & Cross-entropy cost for received pilots and antenna subarray configuration \cellcolor{color1} &Received pilot signals  \cellcolor{color2} & Best antenna subarray index\cellcolor{color1} \\
			\hline
			\hline
			%\hline
			%12 &	$512$ &  & \\
			%\hline 
		\end{tabular}
	\end{table*}
	%------------------------------------------------------------
	%TC:endignore

	To compare the channel characteristics, the accuracy for mmWave-only channels degraded when combined with the V2V channel. This is explained by channel dynamics and the loss of beamforming gain, that mmWave usually leverages upon via multiple antennas. The reliability of ML tasks aggravated in multi-hop scenario, i.e., mmWave-V2V for all delay profiles because of error propagation when the inaccurately received ML data in one vehicle was transmitted to the BS via the mmWave channel. Among all delay profiles, those with nLoS  propagation had the most severe conditions leading to low classification accuracy. In particular, the `Highway nLoS' showed the worst performance while `Rural LoS' fared the best for all tasks in both V2V and mmWave-V2V channels. 
	
	Comparing the classification performance of both ML tasks, higher (\textasciitilde$7\%$) accuracy was obtained for image classification than speech recognition. Note that both had the same number of classes. This is obvious because the image dataset had more distinguishable patterns (e.g., handwritten digits) whereas the features in the spectrogram of the audio signals were less prominent.
	When both training and test datasets were corrupted, the performance improvement in both tasks was within ballpark of each other due to the similarity. That is to say, they both get close to the clean dataset performance (i.e., $98.6\%$ and $91.5\%$ for, respectively, image and speech).

	\section{Challenges  and Future Research Directions}
	%We discuss the design challenges for the reliable implementation of implicit channel learning in various aspects and highlight the possible future research directions.
	The vision of 6G embraces the deployment of ML models for various problems/scenarios at different layers of OSI model, e.g., in the application layer (image recognition)~\cite{survey_DL_Scalable}, physical to transport layer (resource allocation)~\cite{implicitCE_IRS,6GLetaiefJSAC,vehicularFederatedMag}, and physical layer (channel estimation, beamforming)~\cite{deepCNN_ChannelEstimation,vehChanEstDL,elbir2019online}. Thus, it is a challenging task to employ these ML models  for different applications simultaneously. 
	
	\subsection{Learning-related challenges}
	Our implicit channel learning approach is one possible solution to reduce the complexity of ML-based methods by combining the channel estimation and recognition tasks. Likewise, similar multi-task learning techniques may be considered for future work by combining the channel learning and beamforming, localization, and antenna selection as detailed in Table~\ref{tableMultiTasks}.
	
	Unlike prior works on IRS-assisted communications~\cite{implicitCE_IRS} and integrated sensing and communications (ISAC)~\cite{implicitCE_ISAC}, which require \textit{a priori} knowledge  (e.g., user locations, see Table~\ref{tableSummary}), our approach is applicable to such emerging applications. The integration of implicit channel learning for semantic communications is another interesting avenue, wherein the transmission of ``meaning'' of the data is studied in an end-to-end learning manner~\cite{6GLetaiefJSAC}.

	\subsection{Data-related challenges} 
	The performance of implicit channel learning strongly depends on the data used to train the ML model. The training dataset should include various channel conditions to provide a sufficient representation for the wireless channel. %Thus, the dataset collection is a major issue.  
	Collecting the channel data in the field is a tedious process. In order to obtain the channel data under certain channel distributions, generative adversarial networks (GANs) are helpful~\cite{ganGYeLi_conf}. In ML context, GANs are used as generative models  to produce data that follow a certain target distribution. For example, GANs generate synthetic media in ``deepfake" applications, where a person in an image or video is swapped with another person's likeness. In wireless communications, GANs are utilized to synthetically generate various channel conditions. %In order to implicitly learn the wireless channel in ML applications, the usage of GANs  opens new research opportunities for data generation based on both ML data and channel models.
	
	Elimination of channel acquisition in implicit channel learning applies only to the data-types related to ML tasks, which consume a large amount of wireless data traffic~\cite{survey_DL_Scalable}. Thus, decision on performing the channel estimation may be based on the data-type. This information could be accessible from data tags in the physical layer. But then it leads to a resource allocation problem stemming from the assignment of data-type-based channel estimation operations. %, i.e., whether they are ML-related or not.

	%	
	%	
	%		%TC:ignore
	%	%%-----------------------------------------------------
	%	\begin{figure}[t]
	%		\centering
	%		{\includegraphics[draft=false,width=0.9\columnwidth]{GAN.png} } 
	%		\caption{A GAN is composed of a generator and a discriminator, each of which is an artificial neural network. The discriminator attempts to differentiate between the real channel data and the fake one that is generated by the channel generator, which tries to generate plausible data to fool the discriminator to make mistakes. Thus, the GAN learns to generate new data with the same statistics as the training set.
	%		}
	%		%			\vspace*{-5mm}
	%		\label{fig_GAN}
	%	\end{figure}
	%	%%-----------------------------------------------------
	%		%TC:endignore

	\subsection{Communications-related challenges}
	Although 6G enables  the peak data rate of $100$Gb/s, the transmission of training datasets to PS for model training still carries a significant overhead. The solution is 6G-enabled edge intelligence or FL, wherein the ML models are trained at the edge level~\cite{6GLetaiefJSAC,vehicularFederatedMag}. For certain physical layer applications (channel estimation and beamforming), FL has been shown \cite{elbir2021FL4PHY} to provide \textasciitilde$15$ times reduction in the communications overhead over centralized ML (CML) that collects the data from the edge to the PS for training. %Compared to CML, 
	The FL-based architecture also keeps datasets at the edge, which is more privacy-preserving and communications-efficient.  %As a result, FL-based techniques hold the potential to deal with large datasets. 
	Nevertheless, FL-based approaches need further enhancements in terms of computational resources at the edge level. %, which invites future design and research challenges. 
	The possible future solutions may include sparsification, pruning, and quantization of the ML models.

	\section{Summary and Concluding Remarks}

	We introduced  implicit channel learning for ML applications in 6G networks. The proposed method jointly learns the features both in ML data and channel characteristics. By constructing a training dataset under the effect of various channel conditions in different propagation environments, the trained ML model becomes robust against the corruptions/imperfections. %Compared to the conventional systems, where the channel estimation is carried out via model-based techniques, implicit channel learning is advantageous due to renouncing the channel estimation stage while exhibiting satisfactory learning performance. We have concluded the following remarks:
	%~\\
	%${\bullet}$ 
	
	Compared to model-based methods, the ML-based  techniques enhance the estimation performance and robustness against the channel dynamics. This is particularly helpful for highly dynamic channels at mmWave and THz, which also employ extremely large arrays.
	
	%${\bullet}$ 
	%The existing ML-based channel estimation require true channel knowledge as training labels. 
	Our proposed method does not rely on true channel knowledge as training labels and provides an end-to-end learning framework. Thus, unlike existing ML-based channel estimation procedures, it is applicable to other physical layer applications, such as resource allocation, beamforming, and localization for diverse and emerging 6G scenarios, e.g., IRS-assisted networks, ISAC, and holographic communications.
	
	%${\bullet}$ 
	The performance of the ML model relies on the training data collected under various channel conditions, for which the GAN-based methods are helpful to enrich the datasets. 
	%${\bullet}$ 
	FL-based approaches are advantageous in terms of privacy and communications-efficiency. However, training very large learning models via FL could be difficult. %Thus, the model weight pruning techniques can be employed.
	
	%~\\
	%${\bullet}$

	%	The proposed approach is also useful compared to ML-based channel estimation techniques since it does not require the use of ML models in multiple layers of the communications systems. We believe that the implicit channel learning approach paves the way for the development new  ML models to jointly learn the CV data and the wireless channel in the physical and application layers.

	% The performance of the implicit learning approach is tested for image and speech classification tasks under the effect of 5G NR, 4G LTE, and IEEE 802.11p V2V channels for both single-hop (user to BS) and multi-hop (V2V to BS) links. 
	
	%	\lipsum[1]
	
	%	\lipsum[1]
	%	In this work, such as channel estimation~\cite{elbir2020_FL_CE}, resource allocation and beamforming~\cite{elbir2020FL_HB}.
	
	%	\newpage
%	\textcolor{red}{please do not replace the reference file. I have edited it because the references were not correctly formatted. Please use the current file}
	\bibliographystyle{IEEEtran}
	\bibliography{IEEEabrv,references_081}

\begin{thebibliography}{10}
\providecommand{\url}[1]{#1}
\csname url@samestyle\endcsname
\providecommand{\newblock}{\relax}
\providecommand{\bibinfo}[2]{#2}
\providecommand{\BIBentrySTDinterwordspacing}{\spaceskip=0pt\relax}
\providecommand{\BIBentryALTinterwordstretchfactor}{4}
\providecommand{\BIBentryALTinterwordspacing}{\spaceskip=\fontdimen2\font plus
\BIBentryALTinterwordstretchfactor\fontdimen3\font minus
  \fontdimen4\font\relax}
\providecommand{\BIBforeignlanguage}[2]{{%
\expandafter\ifx\csname l@#1\endcsname\relax
\typeout{** WARNING: IEEEtran.bst: No hyphenation pattern has been}%
\typeout{** loaded for the language `#1'. Using the pattern for}%
\typeout{** the default language instead.}%
\else
\language=\csname l@#1\endcsname
\fi
#2}}
\providecommand{\BIBdecl}{\relax}
\BIBdecl

\bibitem{6GLetaiefJSAC}
K.~B. Letaief, Y.~Shi, J.~Lu, and J.~Lu, ``Edge artificial intelligence for
  {6G}: {V}ision, enabling technologies, and applications,'' \emph{IEEE J. Sel.
  Areas Commun.}, vol.~40, no.~1, pp. 5--36, 2021.

\bibitem{6G_AIenabled6G}
H.~Yang, A.~Alphones, Z.~Xiong, D.~Niyato, J.~Zhao, and K.~Wu,
  ``Artificial-intelligence-enabled intelligent {6G} networks,'' \emph{IEEE
  Network}, vol.~34, no.~6, pp. 272--280, 2020.

\bibitem{dl_WCM}
L.~{Dai}, R.~{Jiao}, F.~{Adachi}, H.~V. {Poor}, and L.~{Hanzo}, ``{Deep
  Learning for Wireless Communications: An Emerging Interdisciplinary
  Paradigm},'' \emph{{IEEE} Wireless Commun.}, vol.~27, no.~4, pp. 133--139,
  2020.

\bibitem{mimoDLChannelModelBeamformingFacebook}
A.~Alkhateeb, S.~Alex, P.~Varkey, Y.~Li, Q.~Qu, and D.~Tujkovic, ``Deep
  learning coordinated beamforming for highly-mobile millimeter wave systems,''
  \emph{IEEE Access}, vol.~6, pp. 37\,328--37\,348, 2018.

\bibitem{ganGYeLi_conf}
H.~Ye, G.~Y. Li, B.-H.~F. Juang, and K.~Sivanesan, ``Channel agnostic
  end-to-end learning based communication systems with conditional {GAN},'' in
  \emph{IEEE Global Communications Conference Workshops}, 2018, pp. 1--5.

\bibitem{elbir2019online}
A.~M. Elbir, K.~V. Mishra, M.~R.~B. Shankar, and B.~Ottersten, ``A family of
  deep learning architectures for channel estimation and hybrid beamforming in
  multi-carrier {mm-Wave} massive {MIMO},'' \emph{IEEE Trans. Cognit. Commun.
  Networking}, 2021, in press.

\bibitem{deepCNN_ChannelEstimation}
P.~{Dong}, H.~{Zhang}, G.~Y. {Li}, I.~S. {Gaspar}, and N.~{NaderiAlizadeh},
  ``Deep {CNN}-based channel estimation for {mmWave} massive {MIMO} systems,''
  \emph{{IEEE} J. Sel. Topics Signal Process.}, vol.~13, no.~5, pp. 989--1000,
  2019.

\bibitem{elbir2021FL4PHY}
A.~M. Elbir, A.~K. Papazafeiropoulos, and S.~Chatzinotas, ``Federated learning
  for physical layer design,'' \emph{IEEE Commun. Mag.}, vol.~59, no.~11, pp.
  81--87, 2021.

\bibitem{survey_DL_Scalable}
R.~Mayer and H.-A. Jacobsen, ``Scalable deep learning on distributed
  infrastructures: {C}hallenges, techniques, and tools,'' \emph{ACM Comput.
  Surv.}, vol.~53, no.~1, pp. 1--37, 2020.

\bibitem{vehicularFederatedMag}
J.~Posner, L.~Tseng, M.~Aloqaily, and Y.~Jararweh, ``Federated learning in
  vehicular networks: {O}pportunities and solutions,'' \emph{IEEE Network},
  vol.~35, no.~2, pp. 152--159, 2021.

\bibitem{mimoHybridLeus2}
A.~Alkhateeb, O.~E. Ayach, G.~Leus, and R.~W. Heath, ``Channel estimation and
  hybrid precoding for millimeter wave cellular systems,'' \emph{{IEEE} J. Sel.
  Topics Signal Process.}, vol.~8, no.~5, pp. 831--846, 2014.

\bibitem{80211pChannelEstimation}
Z.~{Zhao}, X.~{Cheng}, M.~{Wen}, B.~{Jiao}, and C.~{Wang}, ``Channel estimation
  schemes for {IEEE} 802.11p standard,'' \emph{{IEEE} Intell. Transp. Syst.
  Mag.}, vol.~5, no.~4, pp. 38--49, 2013.

\bibitem{vehChanEstDL}
A.~K. Gizzini, M.~Chafii, A.~Nimr, and G.~Fettweis, ``Deep learning based
  channel estimation schemes for {IEEE} 802.11p standard,'' \emph{IEEE Access},
  vol.~8, pp. 113\,751--113\,765, 2020.

\bibitem{implicitCE_IRS}
T.~Jiang, H.~V. Cheng, and W.~Yu, ``Learning to reflect and to beamform for
  intelligent reflecting surface with implicit channel estimation,'' \emph{IEEE
  J. Sel. Areas Commun.}, vol.~39, no.~7, pp. 1931--1945, 2021.

\bibitem{implicitCE_ISAC}
W.~Yuan, S.~Li, Z.~Wei, J.~Yuan, and D.~W.~K. Ng, ``Bypassing channel
  estimation for {OTFS} transmission: {A}n integrated sensing and communication
  solution,'' in \emph{IEEE Wireless Communications and Networking Conference
  Workshops}, 2021, pp. 1--5.

\end{thebibliography}
	\balance
	
%	
%	%TC:ignore
%	\begin{IEEEbiographynophoto} {Ahmet M. Elbir} (Senior Member, IEEE) is currently a Research Fellow at Duzce University and the University of Luxembourg. 
%	\end{IEEEbiographynophoto}
%	
%	\begin{IEEEbiographynophoto} {Wei Shi} (Member, IEEE)  is currently a professor at the	School of Information Technology, Carleton 		University, Ottawa, Canada.
%	\end{IEEEbiographynophoto}
%	
%	
%	\begin{IEEEbiographynophoto} {Kumar Vijay Mishra} (Senior Member, IEEE)  is currently,  a National Academies Harry Diamond Distinguished Fellow at the U. S. Army Research Laboratory.
%	\end{IEEEbiographynophoto}
%	\begin{IEEEbiographynophoto} {Anastasios K. Papazafeiropoulos} (Senior Member, IEEE) is currently a Vice-Chancellor Fellow at the University of Hertfordshire and a Research Fellow at the University of Luxembourg. 
%	\end{IEEEbiographynophoto}
%	
%	
%	\begin{IEEEbiographynophoto} {Symeon Chatzinotas} (Senior Member, IEEE) is currently the Head of the research group SIGCOM in the Interdisciplinary Centre for Security, Reliability and Trust, University of Luxembourg.
%	\end{IEEEbiographynophoto}
	%	
	%	%	
	%	%	
	%TC:endignore
	
\end{document}